\documentclass[aps,prd,tightenlines,onecolumn,amsmath,amssymb,showpacs]{revtex4}%
\usepackage{graphicx}
\usepackage{epsfig}
\usepackage{subfigure}
\usepackage{graphics,color}
\usepackage[none]{hyphenat}
\usepackage{hyperref}
\usepackage{amsbsy}
\usepackage{bm}
\newcommand{\be}{\begin{equation}}
\newcommand{\ee}{\end{equation}}
\newcommand{\bea}{\begin{eqnarray}}
\newcommand{\eea}{\end{eqnarray}}
\newcommand{\nn}{\nonumber}
\newcommand{\Tr}{{\mathrm{Tr}}}
\begin{document}

\title{Meson Masses and Mixing Angles in 2+1 Flavor Polyakov Quark Meson Sigma Model
and Symmetry Restoration Effects}
\author{Uma Shankar Gupta}
\email {guptausg@gmail.com}
\author{Vivek Kumar Tiwari}
\email {vivekkrt@gmail.com}
\affiliation{Department of Physics, University of Allahabad, Allahabad 211002, India.}

\date{\today}

\begin{abstract}
The meson masses and mixing angles have been calculated for the scalar
and pseudoscalar sector in the framework of the generalized $2+1$ flavor 
Polyakov loop augmented quark meson linear sigma model. We have given 
the results for two different forms of the effective Polyakov loop 
potential. The comparison of results with the existing calculations in 
the bare $2+1$ quark meson linear sigma model, shows that the restoration
of chiral symmetry becomes sharper due to the influence of the Polyakov 
loop potential. We find that inclusion of the Polyakov loop in quark 
meson linear sigma model together with the  presence of axial anomaly, 
triggers an early and significant melting of the strange condensate. 
We have examined how the inclusion of the Polyakov loop qualitatively 
and quantitatively affects the convergence in the masses of the chiral
partners in pseudoscalar ($\pi$, $\eta$, $\eta'$, $K$) and scalar 
($\sigma$, $a_0$, $f_0$,$\kappa$) meson nonets as the temperature is 
varied on the reduced temperature scale. The role of $U_A(1)$ anomaly
in determining the isoscalar masses and mixing angles for the 
pseudoscalar ($\eta$ and $\eta'$) and scalar ($\sigma$ and $f_0$)
meson complex, has also been investigated in the Polyakov quark 
meson linear sigma model. The interplay of chiral symmetry restoration
effects and the setting up of $U_A(1)$ restoration trend has been 
discussed and analyzed in the framework of the presented model 
calculations. 
\end{abstract}


\maketitle

\section{Introduction}
\label{intr}

The present theoretical understanding of strong interaction physics 
indicates that normal hadronic matter undergoes a phase transition, 
where the individual hadrons dissolve into their constituents and 
produce a collective form of matter known as the Quark Gluon Plasma 
(QGP) under the extreme conditions of high temperature and/or density
\cite{Rischke:03,Muller,Svetitsky}. Relativistic heavy ion collision 
experiments at RHIC (BNL), LHC (CERN) and the future CBM experiments 
at the FAIR facility (GSI-Darmstadt) aim to create and study such a 
collective state of matter.
 
Study of the different aspects of this phase transition, is a tough and
challenging task because Quantum Chromodynamics (QCD), the theory of 
strong interaction, becomes nonperturbative in the low energy limit. 
In the zero quark mass limit, chiral condensate works as an order 
parameter for the spontaneous breakdown of the chiral symmetry in the 
low energy hadronic vacuum of the QCD. For the infinitely heavy quarks, 
in the pure gauge $SU_c(3)$ QCD, the $Z(3)$ (center symmetry of the QCD 
color gauge group) symmetry, which is the symmetry of hadronic vacuum, 
gets spontaneously broken in the high temperature/density regime of QGP. 
Here the expectation value of the Wilson line (Polyakov loop) is related 
to the free energy of a static color charge, hence it serves as the order
parameter of the confinement-deconfinement phase transition 
\cite{Polyakov:78plb}. Even though the center symmetry is always broken 
with the inclusion of dynamical quarks in the system, one can regard the 
Polyakov loop as an approximate order parameter because it is a good 
indicator of the confinement-deconfinement transition 
\cite{Pisarski:00prd,Vkt:06}.
 
We get important information and insights from the lattice QCD calculations
(see e.g.~\cite{Karsch:02,Fodor:03,Allton:05,Aoki:06,Karsch:05,
Karsch:07ax,Cheng:06,Cheng:08}) regarding various aspects of the transition,
like the restoration of chiral symmetry in QCD, order of the 
confinement-deconfinement phase transition, richness of the QCD phase 
structure and mapping of the phase diagram. Since lattice calculations 
are technically involved and various issues are not conclusively settled 
within the lattice community, one resorts to the calculations within the
ambit of phenomenological models developed in terms of effective degrees 
of freedom. These models serve to complement the lattice simulations and 
give much needed insight about the regions of phase diagram inaccessible 
to lattice simulations. A lot of current effective model building activity, 
is centered around combining the features of spontaneous breakdown of both 
chiral symmetry as well as the center $Z(3)$ symmetry of QCD in one single 
model (see for example \cite{Schaefer:07,Digal:01,Kahara:08,Ratti:06,
Ratti:07,Ratti:07npa,Tamal:06,Sasaki:07,Hell:08,
Abuki:08,Ciminale:07,Fu:07,Fukushima:08d77,Fukushima:08d78,Fukushima:09}). 
In these models chiral condensate and Polyakov loop are simultaneously 
coupled to the quark degrees of freedom.

In order to calculate the properties of mesons in hot and dense medium,
several investigations have been done in the two and three flavor
Nambu-Jona-Lasinio (NJL), Polyakov-Nambu-Jona-Lasinio (PNJL) models (e.g.~\cite{Ratti:07prd,Costa:09,Costa:05,Costa:04}) and 
also in the $SU(2)$ version of linear sigma model (e.g.
\cite{Schaefer:07prd,Mocsy:01prc,Bielich:00prl}). Since chiral symmetry
restoration is signaled by parity doubling, these studies look for the
patterns of emerging convergence in the masses of the chiral partners
in pseudoscalar ($\pi$, $\eta$, $\eta'$, $K$) and scalar mesons
($\sigma$, $a_0$, $f_0$, $\kappa$). It is a common knowledge that the 
basic QCD Lagrangian has the global 
$SU_{L+R}(3) \times SU_{L-R}(3) \times U_A(1)$ symmetry. Different 
patterns of spontaneous as well as explicit breaking of 
$SU_{V}(3) \times SU_A(3)$, have been discussed by Lenaghan et al. 
\cite{Rischke:00} in the ambit of $SU(3)$  linear sigma model. 
Schaefer et al. enlarged the linear sigma model with the inclusion 
of quarks \cite{Schaefer:09} and then they studied in the 2+1 flavor 
breaking scenario, the consequences of $SU(3)$ chiral symmetry 
restoration for scalar and pseudoscalar meson masses and mixing angles, 
in the presence as well as the absence of $U_A(1)$ axial symmetry, as 
the temperature is increased through the phase transition temperature. 
The $U_A(1)$ axial symmetry does not exist at the quantum level and as 
shown by 't Hooft \cite{tHooft:76prl}, it gets explicitly broken to 
$Z_{A}(N_f)$ by the instanton effects. The $U_A(1)$ anomaly does not 
let the $\eta'$ meson remain massless Goldstone boson in the chiral 
limit by giving it a mass of about 1 GeV. This happens due to flavor 
mixing, a phenomenon that lifts the degeneracy between the $\pi$ and 
$\eta'$ which otherwise would have been degenerate with $\pi$ in $U(3)$
even if the explicit chiral symmetry breaking is present. There is large 
violation in the Okubo-Zweig-Iizuka rule for both pseudoscalar and scalar 
mesons and ideal mixing is not achieved because of strong flavor mixing 
between nonstrange and strange flavor components of the mesons 
\cite{Costa:09}. Hence $U_A(1)$ restoration will have important observable 
effects on scalar and pseudoscalar meson masses as well as the mixing 
angles.

In a three flavor PNJL model calculation, Costa and collaborators
\cite{Costa:09} have discussed in detail how the inclusion of Polyakov 
loop in the NJL model, affects the results of meson mass and mixing 
angle calculations. However in an earlier paper, they have pointed out 
that the description of the $\eta'$ in the NJL model has some problem 
\cite{Costa:05}. The NJL model does not confine and the meson degrees
of freedom are generated in the model by some prescription. The 
polarization function for the meson gets an imaginary part above the 
$\bar{q}q$ threshold, hence $\eta'$ becomes unbound completely in the 
model soon after the temperature is raised from zero. Thus $\eta'$ in 
the NJL model is not a well defined quantity \cite{fukushima:01}. 
Schaefer et al. \cite{Schaefer:09,Schaefer:08ax} have also made an 
elaborate study of meson masses and mixing angles with and without 
$U_A(1)$ axial anomaly in the 2+1 flavor quark meson linear sigma model 
where the mesons are included in the Lagrangian from the very outset and 
the $U_A(1)$ breaking 't Hooft coupling term is constant. The behavior 
of the scalar and pseudoscalar mixing angles in their calculation is 
opposite to what has been reported in the calculation by Costa et al. 
\cite{Costa:09}. It is worthwhile and important to investigate the 
influence of Polyakov loop on meson mass and mixing angle calculations 
in scalar and pseudoscalar sector, in the framework of generalized 2+1 
flavor quark meson linear sigma model enlarged with the inclusion of 
the Polyakov loop \cite{Schaefer:09ax,Schaefer:09wspax,H.mao09}. Since we 
are lacking in the experimental information on the behavior of mass 
and mixing angle observables in the medium, a comparative study of these 
quantities in different models and circumstances becomes all the more 
desirable. We will be investigating how the inclusion of Polyakov loop, 
qualitatively and quantitatively affects the convergence of the masses 
of chiral partners, when the parity doubling takes place as the 
temperature is increased through $T_c$ and the partial restoration of 
chiral symmetry is achieved. We will also be studying the effect of 
Polyakov loop on the interplay of $SU_A(3)$ chiral symmetry and 
$U_A(1)$ symmetry restoration.

The arrangement of this paper is as follows. In Sec.\ref{sec:model}
we have given the formulation of the model. The description of grand 
potential in the mean field approach has been presented in Sec.
\ref{sec:potmf}. We have derived the modification of meson masses due 
to the $\bar{q}q$ contribution in the presence of Polyakov loop in
Sec.\ref{sec:mixang} where the formulae for meson masses and mixing 
angles have been discussed. In Sec.\ref{sec:eploop}, we will be discussing
the numerical results and plots for understanding and analyzing the 
effect of Polyakov loop on chiral symmetry restoration. Discussion and 
summary is presented in the last Sec.\ref{sec:smry}.

\section{Model Formulation}
\label{sec:model}

We will be working in the generalized three flavor quark meson linear 
sigma model which has been combined with the Polyakov loop potential
\cite{Schaefer:09wspax,H.mao09,Schaefer:09ax}. 
In this model, quarks coming in three flavor are coupled to the 
$SU_V(3) \times SU_A(3)$ symmetric mesonic fields together with 
spatially constant temporal gauge field represented by Polyakov loop 
potential. Polyakov loop field $\Phi(\vec{x})$ is defined as the 
thermal expectation value of color trace of Wilson loop in temporal 
direction 
\be
\Phi = \frac{1}{N_c}\Tr_c L, \qquad \qquad  \Phi^* = \frac{1}{N_c}\Tr_c L^{\dagger}
\ee

where L(x) is a matrix in the fundamental representation of the 
$SU_c(3)$ color gauge group.
\be
\label{eq:Ploop}
L(\vec{x})=\mathcal{P}\mathrm{exp}\left[i\int_0^{\beta}d \tau
A_0(\vec{x},\tau)\right]
\ee
Here $\mathcal{P}$ is path ordering,  $A_0$ is the temporal component
of Euclidean vector field and $\beta = T^{-1}$ \cite{Polyakov:78plb}.
 
The model Lagrangian is written in terms of quarks, mesons, couplings 
and Polyakov loop potential ${\cal U} \left( \Phi, \Phi^*, T \right)$.

\be
\label{eq:Lag}
{\cal L}_{PQMS} = {\cal L}_{QMS} - {\cal U} \big( \Phi , \Phi^* , T \big) 
\ee
where the Lagrangian in quark meson linear sigma model
\bea
\label{eq:Lqms}
{\cal L}_{QMS} =  \bar{q_f} \big( i \gamma^{\mu}D_{\mu}  - g\; T_a\big( \sigma_a 
+ i\gamma_5 \pi_a\big) \big) q_f  + {\cal L}_{m} 
\eea
The coupling of quarks with the uniform temporal background gauge 
field is effected by the following replacement 
$D_{\mu} = \partial_{\mu} -i A_{\mu}$ 
and  $A_{\mu} = \delta_{\mu 0} A_0$ (Polyakov gauge), where 
$A_{\mu} = g_s A^{a}_{\mu} \lambda^{a}/2$. $g_s$ is the $SU_c(3)$ 
gauge coupling. $\lambda_a$ are Gell-Mann matrices in the color 
space, a runs from $1 \cdots 8$. $q_f=(u,d,s)^T$ denotes the quarks 
coming in three flavors and three colors. g is the flavor blind 
Yukawa coupling that couples the three flavor of quarks with nine 
mesons in the scalar ($\sigma_a, J^{P}=0^{+}$) and pseudoscalar 
($\pi_a, J^{P}=0^{-}$) sectors.

The quarks have no intrinsic mass but become massive after 
spontaneous chiral symmetry breaking because of nonvanishing 
vacuum expectation value of the chiral condensate. The mesonic 
part of the Lagrangian has the following form
\bea  
\label{eq:Lagmes}
{\cal L}_{m} & = &\Tr \left( \partial_\mu M^\dagger \partial^\mu
    M \right)
  - m^2 \Tr ( M^\dagger M) -\lambda_1 \left[\Tr (M^\dagger M)\right]^2 \nn \\
  &&  - \lambda_2 \Tr\left(M^\dagger M\right)^2
  +c   \big[det (M) + det (M^\dagger) \big] \nn \\
  && + \Tr\left[H(M + M^\dagger)\right].
\eea
The chiral field M is a $3 \times 3$ complex matrix comprising of 
the nine scalars $\sigma_a$ and the nine pseudoscalar $\pi_a$ mesons.
\be
\label{eq:Mfld}
M = T_a \xi_a= T_a(\sigma_a +i\pi_a)
\ee
Here $T_a$ represent 9 generators of $U(3)$ with 
$T_a = \frac{\lambda_a}{2}$. $a=0,1 \dots ~8$. $\lambda_a$ are 
standard Gell-Mann matrices with $\lambda_0=\sqrt{\frac{2}{3}}\ \bf 1$. 
The generators follow $U(3)$ algebra $\left[T_a, T_b\right]  = if_{abc}T_c$ 
and $\left\lbrace T_a, T_b\right\rbrace  = d_{abc}T_c$ where 
$f_{abc}$ and $d_{abc}$ are standard antisymmetric and symmetric 
structure constants respectively with 
$f_{ab0}=0$ and $d_{ab0}=\sqrt{\frac{2}{3}}\ \bf 1 \ \delta_{ab}$ 
and matrices are normalized as $\Tr(T_a T_b)=\frac{\delta_{ab}}{2}$.

The $SU_L(3) \times SU_R(3)$ chiral symmetry is explicitly broken by 
the explicit symmetry breaking term 
\be
H = T_a h_a
\ee
Here H is a $3 \times 3$ matrix with nine external parameters. 
The $\xi$ field picks up the nonzero vacuum expectation value, 
$\bar{\xi}$ due to the spontaneous breakdown of the chiral symmetry. 
Since $\bar{\xi}$ must have the quantum numbers of the vacuum, 
explicit breakdown of the chiral symmetry is only possible with 
three nonzero parameters $h_0$, $h_3$ and $h_8$. We are neglecting 
isospin symmetry breaking hence we choose $h_0$, $h_8 \neq 0$.
This leads to the $2+1$ flavor symmetry breaking scenario with 
nonzero condensates $\bar{\sigma_0}$ and $\bar{\sigma_8}$.

Apart from $h_0$ and $h_8$, the other parameters in the model are 
five in number. These are the squared tree-level mass of the meson 
fields $m^2$, quartic coupling constants $\lambda_1$ and $\lambda_2$, 
a Yukawa coupling $g$ and a cubic coupling constant $c$ which models 
the $U_A(1)$ axial anomaly of the QCD vacuum.

Since it is broken by the quantum effects, the $U_A(1)$ axial 
which otherwise is a symmetry of the classical Lagrangian, becomes
anomalous~\cite{Weinberg:75} and gives large mass to $\eta'$ meson 
($m_{\eta'} = 940$ MeV). In the absence of $U_A(1)$ anomaly, $\eta'$ 
meson would have been the ninth pseudoscalar Goldstone boson, 
resulting due to the spontaneous break down of the chiral $U_A(3)$ 
symmetry. The entire pseudoscalar nonet corresponding to spontaneously 
broken $U_A(3)$, would consist of the three $\pi$, four $K$, $\eta$ 
and $\eta'$ mesons, which are the massless pure Goldstone
modes when $H = 0$ and they become pseudo Goldstone modes after 
acquiring finite mass due to nonzero $H$ in different symmetry 
breaking scenarios. The particles coming from octet 
($a_0$, $f_0$, $\kappa$) and singlet ($\sigma$) representations 
of $SU_V(3)$ group, constitute scalar nonet 
($\sigma$, $a_0$, $f_0$, $\kappa$). In order to study the chiral 
symmetry restoration at high temperatures, we will be investigating 
the trend of convergence in the masses of chiral partners occurring 
in pseudoscalar ($\pi$, $\eta$, $\eta'$, $K$) and scalar 
($\sigma$, $a_0$, $f_0$, $\kappa$) nonets, in the $2+1$ flavor 
symmetry breaking scenario.

\begin{table*}[!hbt]
\begin{tabular}{|l|l||l|l|} 
\hline \hline
&Scalar Meson Sector && Pseudoscalar Meson Sector \\ \hline
$ m^{2}_{a_{0}}$ & $m^2 +\lambda_1 (x^2 + y^2)+\frac{3 \lambda_2}{2} x^2+ 
\frac{\sqrt{2} c}{2}y $ \qquad &
$ m^{2}_{\pi}$ & $m^2 + \lambda_1 (x^2 + y^2) +
  \frac{\lambda_2}{2} x^2 -\frac{\sqrt{2} c}{2} y$ \\ 
$ m^{2}_{\kappa}$ & $ m^2 + \lambda_1 (x^2 + y^2) + \frac{\lambda_2}{2} (x^2 
+ \sqrt{2} xy + 2 y^2) + \frac{c}{2}x $ \qquad & 
$m^{2}_{K}$ & $m^2 + \lambda_1 (x^2 + y^2) +
  \frac{\lambda_2}{2} (x^2 - \sqrt{2} x y +
    2 y^2) - \frac{c}{2} x$ \\ 
    
$m^2_{s,00}$ &  $m^2 + \frac{\lambda_1}{3} (7 x^2 + 4 \sqrt{2} xy + 5 y^2) + 
\lambda_2(x^2 + y^2) - \frac{\sqrt{2}c}{3} (\sqrt{2} x +y)$ & 
$m^2_{p,00}$ & $m^2 + \lambda_1(x^2 +y^2) + \frac{\lambda_2}{3}(x^2 +y^2) + \frac{c}{3} (2x + \sqrt{2} y)$ \\
$m^2_{s,88}$ &  $m^2 + \frac{\lambda_1}{3} (5x^2 -4 \sqrt{2} xy +7y^2) + 
\lambda_2(\frac{x^2}{2} +2y^2) + \frac{\sqrt{2}c}{3} (\sqrt{2}x - \frac{y}{2})$ & 
$m^2_{p,88}$ &  $m^2 +\lambda_1(x^2 +y^2) +\frac{\lambda_2}{6}(x^2 +4y^2) 
-\frac{c}{6}(4x -\sqrt{2}y)$ \\
$m^2_{s,08}$ &  $\frac{2\lambda_1}{3}(\sqrt{2}x^2 -xy -\sqrt{2}y^2) +
\sqrt{2}\lambda_2(\frac{x^2}{2}-y^2) +\frac{c}{3\sqrt{2}}(x- \sqrt{2}y)$ & 
$m^2_{p,08}$ &  $\frac{\sqrt{2}\lambda_2}{6} (x^2 -2y^2) -\frac{c}{6}(\sqrt{2}x -2y)$\\

$ m^{2}_{\sigma} $& $m^2_{s,{00}} \cos^{2}\theta_s + m^2_{s,88}\sin^{2}\theta_s + 2  m^{2}_{s,08} \sin \theta_s \cos
 \theta_s$ \qquad & 
$m^{2}_{\eta'}$ & $m^2_{p,00} \cos^{2}\theta_p + m^2_{p,88}\sin^{2}\theta_p + 2 m^{2}_{p,08} \sin \theta_p \cos \theta_p$ \\ 
$m^{2}_{f_0}$  & $m^2_{s,00} \sin^{2}\theta_s +
 m^{2}_{s,88}\cos^{2}\theta_s - 2  m^2_{s,08} \sin \theta_s \cos \theta_s $ \qquad &
 $ m^{2}_{\eta}$ & $m^2_{p,00} \sin^{2}\theta_p + m^2_{p,88}\cos^{2}\theta_p - 2
 m^{2}_{p,08} \sin \theta_p \cos \theta_p$ \\ 

$m^{2}_{\sigma_{NS}}$ &  $\frac{1}{3}(2m^{2}_{s,00} +m^{2}_{s,88} +2\sqrt{2}m^{2}_{s,08})$&
$m^{2}_{\eta_{NS}}$ &  $\frac{1}{3}(2m^{2}_{p,00} +m^{2}_{p,88} +2\sqrt{2}m^{2}_{p,08})$\\
$m^{2}_{\sigma_{S}}$ &  $\frac{1}{3}(m^{2}_{s,00} +2m^{2}_{s,88} -2\sqrt{2}m^{2}_{s,08})$&
$m^{2}_{\eta_{S}}$ &  $\frac{1}{3}(m^{2}_{p,00} +2m^{2}_{p,88} -2\sqrt{2}m^{2}_{p,08})$\\
  \hline
\end{tabular}
 \label{tab:mass}
\caption{The squared masses of scalar and pseudoscalar mesons appear 
in nonstrange-strange basis. In the following table x denotes $\sigma_x$ 
and y denotes $\sigma_y$. The  masses of nonstrange $\sigma_{NS}$, 
strange $\sigma_S$, nonstrange $\eta_{NS}$ and strange $\eta_S$  mesons 
are given in the last two rows.}
  \end{table*}

\subsection{Choice of Potentials for the Polyakov Loop}

The effective potential ${\cal U} \left( \Phi, \Phi^*, T \right)$ 
is constructed such that it reproduces thermodynamics of pure glue 
theory on the lattice for temperatures upto about twice the 
deconfinement phase transition temperature. At much higher temperatures, 
the transverse gluons become effective degrees of freedom, hence the 
construction of effective potential in terms of the Polyakov loop 
potential is not reliable \cite{Ratti:06,Ratti:07prd}.

At low temperatures, the effective potential  
${\cal U} \left( \Phi, \Phi^*, T \right)$ has  only one minimum at 
$\Phi = 0$ in the confined phase. Above the critical temperature 
for deconfinement transition, $\Phi = 0$ becomes metastable local 
minimum and now, the effective potential has three degenerate 
global minima at $\Phi \neq 0$ due to the spontaneous breakdown of 
the $Z(3)$ center symmetry.

In this work, we use the following two choices of the effective 
potential. The first choice is based on the polynomial expansion 
in terms of Polyakov loop order parameter $\Phi$ and is given 
\cite{Ratti:06} as
\bea 
\label{eq:polpot}
\frac{{\cal U_{\text{pol}}}\left(\Phi,\Phi^*, T \right)}{T^4} &=& -\frac{b_2}{4}
\left(|\Phi|^2+|{\Phi^{*}|}^{2} \right) -\frac{b_3}{6}(\Phi^3+{\Phi^*}^3) \nn \\
&&+\frac{b_4}{16} \left(|\Phi|^2+|\Phi^*|^2\right)^2
\eea 
The second term that is the sum of $\Phi^3$ and $\Phi^{*3}$ terms, 
causes the three degenerate vacua above the deconfinement phase 
transition. The potential parameters are adjusted according to the 
pure gauge lattice data such that the equation of state and Polyakov 
loop expectation values are reproduced. The temperature dependent 
coefficient $b_2(T)$ governs the confinement-deconfinement phase 
transition and is given by 
\begin{equation*}
b_2(T) = a_0 + a_1 \left(\frac{T_0}{T}\right) + a_2
\left(\frac{T_0}{T}\right)^2 \nn \\+ a_3 \left(\frac{T_0}{T}\right)^3\ . 
\end{equation*}

The other parameters have the following value
\begin{eqnarray*}
&& a_0=6.75\ , \quad a_1=-1.95\ , \quad a_2=2.625\ ,\nn \\
&& a_3 = -7.44,  \quad  b_3=0.75\ , \quad b_4=7.5\ . 
\end{eqnarray*}

The other choice of effective potential as given in 
Ref.~\cite{Ratti:07}, has the logarithmic form. The results 
produced by this potential are known to be fitted well to 
lattice results.
\bea
\label{eq:logpot}
\frac{{\cal U_{\text{log}}}\left(\Phi,\Phi^*, T \right)}{T^4} &=& -\frac{a\left(T\right)}{2}\Phi^* \Phi +
    b(T) \, \mbox{ln}[1-6\Phi^* \Phi \nn \\
    &&+4(\Phi^{*3}+ \Phi^3)-3(\Phi^* \Phi)^2]
\eea

where the temperature dependent coefficients are as follow

\begin{equation*} 
  a(T) =  a_0 + a_1 \left(\frac{T_0}{T}\right) + a_2 \left(\frac{T_0}{T}\right)^2 \; \; \; 
  b(T) = b_3 \left(\frac{T_0}{T}\right)^3\ .
\end{equation*}

The critical temperature for deconfinement phase transition 
$T_0=270$ MeV is fixed for pure gauge sector. The parameters 
of Eq.(\ref{eq:logpot}) are 
\begin{eqnarray*}
&& a_0 = 3.51\ , \qquad a_1= -2.47\ , \nn \\ 
&& a_2 = 15.2\ ,  \qquad  b_3=-1.75\ 
\end{eqnarray*}

Both effective potential fits reproduce equally well the equation
of state and the Polyakov loop expectation value.

\begin{table*}[!ht]
 \begin{tabular}{|c||c|c|c|c|c|c|}
\hline \hline
& C[MeV] & $m^2$ $[MeV^2]$  & $\lambda_1$ & $\lambda_2$ & $h_x$ $[MeV^3]$& $h_y$ $[MeV^3]$ \\
\hline
W/$U_A(1)$&    4807.84      &   $(342.52)^2$    &   1.40      & 46.48 & $(120.73)^3$ &$(336.41)^3$\\
W/o$U_A(1)$&        0       &$-(189.85)^2$&  -17.01     & 82.47 & $(120.73)^3$ & $(336.41)^3$ \\ \hline 
 \end{tabular}
\label{tab:pmtr}
\caption{parameters for $m_{\sigma} = 600$ MeV with and without $U_A(1)$
axial anomaly term.}
\end{table*}
  
\section{Grand Potential in the Mean-Field Approach}
\label{sec:potmf}

The thermodynamics of changing numbers of particles and antiparticles 
is governed by grand canonical partition function. We are considering 
a spatially uniform system in thermal equilibrium at finite temperature
T and quark chemical potential $\mu_f (f=u, d, s)$. The partition 
function is written as the path integral over quark/antiquark and meson 
fields \cite{Schaefer:09}
\bea
\label{eq:partf}
\mathcal{Z}&=& \mathrm{Tr\, exp}[-\beta (\hat{\mathcal{H}}-\sum_{f=u,d,s} 
\mu_f \hat{\mathcal{N}}_f)] \nn \\
&=& \int\prod_a \mathcal{D} \sigma_a \mathcal{D} \pi_a \int
\mathcal{D}q \mathcal{D} \bar{q} \; \mathrm{exp} \bigg[- \int_0^{\beta}d\tau\int_Vd^3x  \nn \\
&& \bigg(\mathcal{L_{QMS}^{E}} 
 + \sum_{f=u,d,s} \mu_{f} \bar{q}_{f} \gamma^0 q_{f} \bigg) \bigg]. 
\eea

where V is the three dimensional volume of the system, and 
$\beta= \frac{1}{T}$. For three quark flavors, in general, the 
three quark chemical potential are different. In this work, we 
assume that $SU_V(2)$ symmetry is preserved and neglect the small 
difference in masses of u and d quarks. Thus the quark chemical 
potential for u and d quarks become equal $\mu_x = \mu_u = \mu_d$.
The strange quark chemical potential is $\mu_y = \mu_s$. Further we 
consider symmetric quark matter and net baryon number to be zero.

The partition function for the $SU(3)$ version of the linear sigma 
model with or without quarks can be evaluated by the more advanced 
many-body resummation techniques such as the self consistent Hartree 
approximation in the Cornwall, Jackiw and Tomboulis 
\cite{CJT,Rischke:00} formalism or the so called optimized perturbation 
theory \cite{Hatsuda:98} in its improved version
\cite{Herpay:05,Herpay:06,Herpay:07}. However the predictive power of 
these methods depends on how they are implemented in different 
approximation schemes. 

In the simple mean field approximation, one does not encounter 
various problems of the more advanced many-body resummation 
techniques. In our work, the partition function has been evaluated 
in the mean-field approximation 
\cite{Mocsy:01prc,Schaefer:08ax,Schaefer:09}. We replace meson field 
by their expectation values 
$\langle \Phi \rangle =  T_0 \bar{\sigma_0} + T_8 \bar{\sigma_8}$ 
and neglect both thermal as well as quantum fluctuations of meson 
fields while quarks and antiquarks are retained as quantum field. 
Now following the standard procedure as given in 
Refs.~\cite{Kapusta_Gale,Schaefer:07,Ratti:06,Fukushima:04plb}
one can obtain the expression of grand potential as sum of pure 
gauge field contribution ${\cal U} \left(\Phi, \Phi^*, T \right)$, 
meson contribution and quark/antiquark contribution evaluated in 
the presence of the Polyakov loop,

\bea
\label{eq:grandp}
\Omega (T, \mu ) = - \frac{T\ln Z}{V} &=& U(\sigma_0, \sigma_8) + {\cal U} \left(\Phi, \Phi^*, T \right) \nn \\
&&+ \Omega_{\bar{q}q} (T, \mu)
\eea 

In order to study 2 + 1 flavor case, one performs following 
basis transformation of condensates and external fields from 
original singlet octet (0, 8) basis to nonstrange strange basis 
(x, y).
\bea
\sigma_x &=&
\sqrt{\frac{2}{3}}\bar{\sigma}_0 +\frac{1}{\sqrt{3}}\bar{\sigma}_8, \\
\sigma_y &=&
\frac{1}{\sqrt{3}}\bar{\sigma}_0-\sqrt{\frac{2}{3}}\bar{\sigma}_8.
\eea

Similar expressions exist for writing the external fields 
($h_x$, $h_y$) in terms of ($h_0$, $h_8$). Thus the nonstrange 
and strange quark/antiquark decouple and the quark masses become
\be 
m_x = g \frac{\sigma_x}{2}, \qquad m_y = g \frac{\sigma_y}{\sqrt{2}}
\ee 
Quarks become massive in symmetry broken phase because of non 
zero vacuum expectation values of the condensates.

The mesonic potential in the nonstrange-strange basis reads,
\bea
\label{eq:mesop}
 U(\sigma_{x},\sigma_{y}) &= &\frac{m^{2}}{2}\left(\sigma_{x}^{2} +
  \sigma_{y}^{2}\right) -h_{x} \sigma_{x} -h_{y} \sigma_{y}
 - \frac{c}{2 \sqrt{2}} \sigma_{x}^2 \sigma_{y} \nn \\
 && + \frac{\lambda_{1}}{2} \sigma_{x}^{2} \sigma_{y}^{2}+
  \frac{1}{8}\left(2 \lambda_{1} +
    \lambda_{2}\right)\sigma_{x}^{4} \nn \\
 && +\frac{1}{8}\left(2 \lambda_{1} +
    2\lambda_{2}\right) \sigma_{y}^{4}\ ,
\eea    

The chiral part of the Polyakov loop augented quark meson linear 
sigma (PQMS) model has the six input parameters and 
therefore require six known quantities as input. In general 
$m_{\pi}$, $m_K$, the pion and kaon decay constant $f_\pi$, $f_K$,
mass square of $\eta$, $\eta'$ and $m_\sigma$ are used to fix 
these parameters. The parameters are fitted such that in vacuum 
the model produces observed pion mass 138 MeV. In the present 
work we are using the set of parameters for sigma mass 
$m_{\sigma} = 600$ MeV. The parameters used in this work, taken 
from \cite{Schaefer:09}, are shown in Table \ref{tab:pmtr}. 

Finally the quark/antiquark Polyakov loop contribution reads,
\bea 
\Omega_{\bar{q}q} (T, \mu) = -2T \sum_{f=u,d,s} \int \frac{d^3 p}{(2\pi)^3}
\Bigl[ \ln g_{f}^{+} + \ln g_{f}^{-} \Bigr]
\eea 

We define $g_{f}^{+}$ and $g_{f}^{-}$ after taking trace over 
color space
\bea
\label{eq:gpls} 
  g_{f}^{+} =  \Big[ 1 + 3\Phi e^{ -E_{f}^{+} /T} +3 \Phi^*e^{-2 E_{f}^{+}/T} +e^{-3 E_{f}^{+} /T}\Big] \,
\eea
\bea
\label{eq:gmns} 
  g_{f}^{-} =  \Big[ 1 + 3\Phi^* e^{ -E_{f}^{-} /T} +3 \Phi e^{-2 E_{f}^{-}/T} +e^{-3 E_{f}^{-} /T}\Big] \,
\eea

Here we use the notation E$_{f}^{\pm} =E_f \mp \mu $ and $E_f$ is the
flavor dependent single particle energy of quark/antiquark.
\be 
E_f = \sqrt{p^2 + m{_f}{^2}}
\ee 

$m{_f}$ is flavor dependent quark mass and is function of 
condensates $\sigma_0$ and $\sigma_8$.

One can very easily notice from equations (\ref{eq:gpls}) and 
(\ref{eq:gmns}) that the role of quarks and antiquarks as well as 
that of the Polyakov loop and its conjugate can be interchanged by the 
transformation $\mu \rightarrow -\mu$. Confinement is the very 
interesting feature of the QCD and the PQMS model describes this 
behavior qualitatively. The Polyakov loop is order parameter for
confinement-deconfinement phase transition. In the confined phase 
$\Phi = 0$. It can also be noticed from the grand potential that one 
and two quark state contributions are vanishing. Only the three 
quark states contribute. In this way the PQMS model qualitatively mimics 
confinement of quark/antiquark within three quark color singlet 
states \cite{Sasaki:07}.

One can get the quark condensates $\sigma_x$, $\sigma_y$ and the Polyakov
loop expectation values $\Phi$, $\Phi^*$ by searching the global 
minima of the grand potential for a given value of temperature T 
and chemical potential 
$\mu$.

\be 
\label{eq:gapeq}
  \left.\frac{ \partial \Omega}{\partial
      \sigma_x} = \frac{ \partial \Omega}{\partial \sigma_y} 
      = \frac{ \partial \Omega}{\partial \Phi}
      = \frac{\partial \Omega}{\partial \Phi^*}
  \right|_{\sigma_x = \bar\sigma_x, \sigma_y=\bar\sigma_y, 
   \Phi=\bar\Phi, \Phi^* =\bar\Phi^*} = 0\ .
\ee 

In this work we are always considering the $\mu = 0$ case.

\begin{table}[!b]
\begin{tabular}{|cc|cccc|}
\hline \hline
&& $m^{2}_{x,a} m^{2}_{x,b}/g^4$ & $m^{2}_{x,ab}/g^2$& $m^{2}_{y,a} m^{2}_{y,b}/g^4$ & $m^{2}_{y,ab}/g^2$\\
\hline
$\sigma_0$ & $\sigma_0$ &$\frac{1}{3}\sigma_{x}^{2}$& $\frac{2}{3}$& $\frac{1}{3}
\sigma_{y}^{2}$& $\frac{1}{3}$ \\
$\sigma_1$ & $\sigma_1$ &$\frac{1}{2}\sigma_{x}^{2}$& $ 1$ & $0$ & $0$\\
$\sigma_4$ & $\sigma_4$ &$ 0$ &$\sigma_x \frac{\sigma_x + \sqrt{2} \sigma_y}
{\sigma_{x}^{2} -2 \sigma_{y}^{2}}$ & $0$ & $\sigma_y \frac{ \sqrt{2} \sigma_x +2 \sigma_y}{2 \sigma_{y}^{2} - \sigma_{x}^{2}}$ \\
$\sigma_8$ & $\sigma_8$ &$ \frac{1}{6} \sigma_{x}^{2}$ & $\frac{1}{3}$ & 
$ \frac{2}{3}\sigma_{y}^{2} $ & $\frac{2}{3} $ \\
$\sigma_0$ & $\sigma_8$ & $ \frac{\sqrt{2} }{6} \sigma_{x}^{2}$ & $ \frac{\sqrt{2}}{3}$ & 
$- \frac{\sqrt{2}}{3} \sigma_{y}^{2}$ & $- \frac{\sqrt{2}}{3}$\\ 
$\pi_0$ & $\pi_0$ & $0$ & $\frac{2}{3}$ & $0$ & $\frac{1}{3}$ \\
$\pi_1$ & $\pi_1$ & $0$ & $1$ & $0$ & $0$ \\
$\pi_4$ & $\pi_4$ & $0$ & $\sigma_x \frac{\sigma_x -\sqrt{2}\sigma_y}{\sigma_{x}^{2} 
- 2 \sigma_{y}^{2}}$ & $0$ & $\sigma_y \frac{\sqrt{2} \sigma_x -2 \sigma_y}
{\sigma_{x}^{2} -2 \sigma_{y}^{2}}$ \\
$\pi_8$ & $\pi_8$ & $0$ & $\frac{1}{3}$ & $0$ & $\frac{2}{3}$\\
$\pi_0$ & $\pi_8$ & $0$ & $\frac{\sqrt{2}}{3}$ & $0$ & $-\frac{\sqrt{2}}{3}$ \\ 
\hline
 \end{tabular}
 \label{tab:qmassd}
\caption{First and second derivative of squared quark mass in 
nonstrange-strange basis with respect to meson fields are evaluated 
at minimum. Sum over two light flavors, denoted by symbol x, are in 
third and fourth columns. The last two columns have only strange quark 
mass flavor denoted by the symbol y.}
 \end{table}

\section{Meson masses and Mixing angles}
\label{sec:mixang}

The curvature of grand potential Eq.(\ref{eq:grandp}) at the global 
minimum determines scalar and pseudoscalar meson masses.
\be 
\label{eq:sdergrand}
 m_{\alpha,ab}^{2} = \frac{\partial^2 \Omega (T, \mu)}{\partial \xi_{\alpha,a} 
 \partial \xi_{\alpha,b}} \bigg|_{min}
\ee 
where subscript $\alpha=$ s, p; s stands for scalar and p stands 
for pseudoscalar meson and a, $b=0 \cdots 8$. We note that the 
Polyakov loop decouples from the mesonic sector at T=0 and the 
meson masses do not receive contribution from quark/antiquark in 
vacuum and hence meson masses are governed by mesonic potential 
only. The mesonic contribution to the meson masses is summarized 
in Table \ref{tab:mass}. The diagonalization of (0 - 8) component 
of mass matrix gives masses of $\sigma$ and $f_0$ mesons in scalar 
sector and masses of $\eta'$ and $\eta$ in pseudoscalar sector. 
The scalar mixing angle $\theta_s$ and pseudoscalar mixing angle 
$\theta_p$ are given by,
\be 
\tan{2\theta_{\alpha}} = \biggl( \frac{2 m^2_{\alpha,08}}
{m^2_{\alpha,00}-m^2_{\alpha,88}} \biggr)
\ee
Here $\alpha$ stands for scalar and pseudoscalar field. The detail 
expressions for masses and mixing angles are given in 
Ref.~\cite{Rischke:00,Schaefer:09}. The meson masses are further 
modified in medium at finite temperature by the quark contributions 
in the grand potential. In order to calculate the second derivative
Eq.(\ref{eq:sdergrand}) for evaluating the quark contribution in the 
presence of the Polyakov loop potential, the complete dependence of all 
scalar and pseudoscalar meson fields Eq.(\ref{eq:Mfld}) has to be 
taken into account. We have to diagonalize the resulting quark mass 
matrix. The expression for the meson mass modification due to quark 
contribution at finite temperature in QMS model, has been evaluated 
by Schaefer et al. \cite{Schaefer:09} and is given as
\bea
\label{eq:ftlsmass}
 \delta m_{\alpha,ab}^{2} &=& \frac{\partial^2 \Omega_{\bar{q}q} (T, \mu)}
 {\partial \xi_{\alpha,a} \partial \xi_{\alpha,b}} \Big|_{min} = \nu_{c} \sum_{f=x,y} 
 \int \frac{d^3 p}{(2\pi)^3} \frac{1}{2E_f} \nn \\
&& \biggl[ (a^{+}_{f} + a^{-}_{f} ) \biggl( m^{2}_{f,a b} - \frac{m^{2}_{f,a}
 m^{2}_{f, b}}{2 E_{f}^{2}} \biggl)  \nn \\
&& - (b_{f}^{+} + b_{f}^{-}) \biggl(  \frac{m^{2}_{f,a}  m^{2}_{f, b}}{2 E_{f} T}
\biggl) \biggl]  
\eea 

$m^{2}_{f,a} \equiv \partial m^{2}_{f}/ 
\partial \xi_{\alpha,a} $ is the first derivative and 
$m^{2}_{f,ab} \equiv \partial m^{2}_{f,a}/ \partial \xi_{\alpha,b}$ 
is the second derivative of squared quark mass with respect to meson 
fields $\xi_{\alpha,b}$. The number of internal quark degrees of 
freedom, $\nu_{c} = 2N_{c} = 6$. Here $a^{\pm}_{f}$ are 
quark/antiquark occupation numbers; given as
\be 
a^{\pm}_{f} = \frac{1}{1+ e^{E^{\pm}/T}}
\ee 
and the notations $b^{\pm}_{f} = a^{\pm}_{f} - (a^{\pm}_{f})^2$ 
stand for particle $(+)$ and antiparticle $(-)$ in quark meson 
linear sigma model without inclusion of the Polyakov loop.

The expression of mass modification due to quark contribution 
at finite temperature, will change in the presence of the Polyakov 
loop. We are obtaining the following formula for the mass 
modification that results on account of quark contribution in the 
PQMS model
\bea
\label{eq:ftmass}
 \delta m_{\alpha,ab}^{2} &=& \frac{\partial^2 \Omega_{\bar{q}q} (T, \mu)}
 {\partial \xi_{\alpha,a} \partial \xi_{\alpha,b}} \Big|_{min} = 3 \sum_{f=x,y} 
 \int \frac{d^3 p}{(2\pi)^3} \frac{1}{E_f} \nn \\
&& \biggl[ (A^{+}_{f} + A^{-}_{f} ) \biggl( m^{2}_{f,a b} - \frac{m^{2}_{f,a}
 m^{2}_{f, b}}{2 E_{f}^{2}} \biggl)  \nn \\
&& + (B_{f}^{+} + B_{f}^{-}) \biggl(  \frac{m^{2}_{f,a}  m^{2}_{f, b}}{2 E_{f} T}
\biggl) \biggl]  
\eea 

The notations $A_{f}^{\pm}$ and $B_{f}^{\pm}$ have the following 
definitions
\be
A_{f}^{+} =  \frac{\Phi e^{-E_{f}^{+}/T} + 2 \Phi^* e^{-2E_{f}^{+}/T} + e^{-3E_{f}^{+}/T}}{g_{f}^{+}}
\ee

\be
A_{f}^{-} =  \frac{\Phi^* e^{-E_{f}^{-}/T} + 2 \Phi e^{-2E_{f}^{-}/T} + e^{-3E_{f}^{-}/T}}{g_{f}^{-}}
\ee

and $B_{f}^{\pm}={3(A_{f}^{\pm}})^2 - C_{f}^{\pm}$, where we 
again define 

\be
C_{f}^{+} = \frac{\Phi e^{-E_{f}^{+}/T} + 4 \Phi^* e^{-2E_{f}^{+}/T} +3 e^{-3E_{f}^{+}/T}}{g_{f}^{+}}
\ee

\be
C_{f}^{-} = \frac{\Phi^* e^{-E_{f}^{-}/T} + 4 \Phi e^{-2E_{f}^{-}/T} +3 e^{-3E_{f}^{-}/T}}{g_{f}^{-}}
\ee

The squared quark mass derivatives evaluated at minimum which were 
originally derived in Ref.\cite{Schaefer:09}, are collected in 
Table \ref{tab:qmassd}. The inclusion of Polyakov loop in QMS model 
does not make any change in these equations. 
%
%

\section{Effect of The Polyakov Loop on The Restoration of Chiral Symmetry }
\label{sec:eploop}

\begin{table}[!b]
\begin{tabular}{|c|c|c|c|}
\hline 
& QMS & PQMS:pol & PQMS:log    \\
\hline
$T_{c}^{\chi}$ (MeV) & $146 $ & $204$ & $206$ \\
$T_{s}^{\chi}$ (MeV) & $248 $ & $262$ & $274$ \\
$T_{c}^{\Phi}$ (MeV) & $ -  $ & $204$ & $206$ \\
\hline
 \end{tabular}
  \label{tab:ctemp}
\caption{The characteristic temperature (pseudocritical temperature) 
for the chiral transition in the nonstrange sector $T_{c}^{\chi}$, 
strange sector $T_{s}^{\chi}$ and confinement-deconfinement 
transition $T_{c}^{\Phi}$, in the QMS, PQMS:log and PQMS:pol models.}
 \end{table}

\begin{figure*}[!t]
\centering
\hspace{-0.3cm}\includegraphics[width=0.8\linewidth]{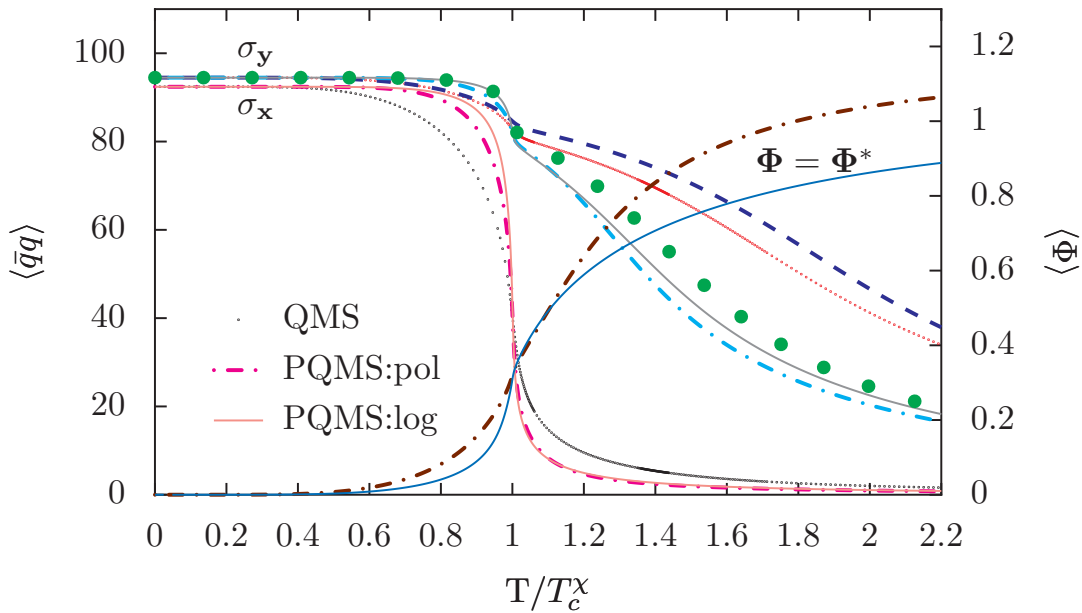}
\caption{The variation of nonstrange $\sigma_x$, strange $\sigma_y$ 
condensates with respect to the relative temperature scale 
($T/T_{c}^{\chi}$) at zero chemical potential ($\mu = 0$) in the QMS 
model and PQMS models with polynomial and logarithmic potentials for 
the Polyakov loop is shown. The lines with continuous dots represent 
the variation in the QMS model, while the dashed-dotted lines show the variation 
in PQMS:pol model and the solid lines are the variations in the
PQMS:log model. The line with the big solid dots shows the $\sigma_y$ 
variation in the PQMS:log model, while the dark dashed line shows the 
pure QMS model results when anomaly is absent, i.e. $c=0$. The 
expectation value of the Polyakov loop $\langle\Phi\rangle$, in PQMS:pol 
and PQMS:log model is shown in the right plots.}
\label{fig:fig1}
\end{figure*}

%
We are presenting the result of our calculation for estimating 
the effect of the Polyakov loop potential on the restoration of chiral 
symmetry when it is included in the $2+1$ flavor quark meson linear
sigma model at finite temperature and zero chemical potential 
with and without axial $U_A(1)$ breaking. We have considered the two 
different ansatzs for the Polyakov loop potential namely the polynomial 
potential and logarithmic potential and compared the results with 
the existing calculations in the quark meson linear sigma model
\cite{Schaefer:09}. The interplay of the effect of $U_A(1)$ axial 
restoration and chiral symmetry restoration in the presence of the
Polyakov loop potential has been shown through the temperature 
variation of strange, nonstrange chiral condensates, meson masses 
and mixing angles. The $U_A(1)$ axial breaking term has been kept 
constant throughout the investigation. The value of Yukawa coupling 
g has been fixed from the nonstrange constituent quark mass 
$m_q = 300$ MeV and is equal to 6.5. This predicts the strange quark 
mass $m_s \backsimeq 433$ MeV.
\subsection{ Condensates and the Polyakov Loop}

The solutions of the coupled gap equations, Eq.(\ref{eq:gapeq}) 
determine the nature of chiral and deconfinement phase transition 
through the temperature and chemical potential dependence of 
nonstrange and strange condensates ($\sigma_x$ and $\sigma_y$) 
and the expectation value of the Polyakov loop ($\langle\Phi\rangle$ 
and $\langle\Phi^*\rangle$). The temperature variation of $\sigma_x$,
$\sigma_y$ and $\langle\Phi\rangle$ in mean-field approximation, 
at zero chemical potential in the PQMS models
with the polynomial Polyakov loop potential
Eq.(\ref{eq:polpot}) (PQMS:pol) and the logarithmic Polyakov loop
potential Eq.(\ref{eq:logpot}) (PQMS:log) is shown in Fig\ref{fig:fig1}.
We have also plotted the strange and nonstrange condensate in QMS
model to compare and investigate the effect of the Polyakov loop
potential inclusion
on chiral symmetry restoration trend reflected through masses
and mixing angles of mesonic excitations. The characteristic
temperatures (pseudocritical temperature) for the confinement -
deconfinement transition $T_{c}^{\Phi}$, the chiral transition
in the nonstrange sector $T_{c}^{\chi}$ and strange sector 
$T_{s}^{\chi}$ are defined through the
inflection point of $\langle\Phi\rangle$, $\sigma_x$, and $\sigma_y$.
We note that the $\langle\Phi\rangle$  = $\langle\Phi^*\rangle$ at 
zero chemical potential. The numerical value of
the pseudocritical temperature for various transitions in the 
QMS model and the PQMS model with the
polynomial and the logarithmic potentials for the Polyakov loop has been 
given in Table \ref{tab:ctemp}. It is evident from the table that the 
chiral transition gets shifted to the higher temperatures as a result 
of the inclusion of the Polyakov loop potential in the QMS model.

We have chosen to compare the results of our calculation in the 
PQMS model with the corresponding results in the QMS model on a
relative temperature scale $T/T_{c}^{\chi}$. Such a choice is 
justified on account of the Ginzburg-Landau effective theory, 
where absolute comparison of the characteristic temperatures 
between two models of the same universality class can not be 
made \cite{Costa:09}.
The condensates start with fixed values $\sigma_x = 92.4$ MeV and 
$\sigma_y = 94.5$ MeV at $T = 0$ as shown in Fig\ref{fig:fig1}.
It is known from the lattice simulations
that transition from hadronic matter to quark gluon 
plasma is a analytic and rapid crossover (\cite{Karsch:02,Aoki:06}).
The Polyakov loop potential inclusion in the QMS models makes the
crossover in $SU_L(2) \times SU_R(2)$ sector quite sharp as the 
nonstrange condensate $\sigma_x$ changes rapidly in the 
transition region. The $U_A(1)$ anomaly does not cause any difference in 
the behavior of nonstrange condensate and $\sigma_x$ remain unchanged
in the presence as well as in the absence of $U_A(1)$ anomaly term. The 
variation of the strange condensate is lot more smooth on account of 
the large constituent mass of the strange quark $m_s$ = 433 MeV.
The Polyakov loop potential inclusion has a strong effect 
on the strange condensate variation also and generates a significant 
melting of $\sigma_y$ in our calculation. 
The interesting physical consequences of the earlier and 
significant melting of the strange condensate will be an early 
emergence of mass degeneration trend in the masses of the chiral 
partners ($K$, $\kappa$) and ($\eta$, $f_0$) and an early setting 
up of a $U_A(1)$ restoration trend on reduced temperature scale.
In the presence of the $U_A(1)$ anomaly, $\sigma_y$ temperature 
variation shows a little more decrease in the respective cases.

Curves starting from the right end of the plot represent 
the variation of the Polyakov loop expectation value 
$\langle\Phi\rangle$ on the relative temperature scale at 
zero chemical potential. 
Though the thermodynamics of quark gluon plasma reproduced with 
the Polyakov loop polynomial potential is found to be in 
agreement with that of lattice simulations, upto twice of the
critical temperature (\cite{Ratti:06,Schaefer:07,Schaefer:09ax})
at higher temperatures $\langle\Phi\rangle$ increases
above unity and this is unphysical.
In the improved ansatz, logarithmic
potential replaces the higher order terms of $\Phi$ and $\Phi^*$ 
in the polynomial potential by the logarithm of Jacobi determinant
which results from integrating out six nondiagonal Lie algebra
directions while keeping the two diagonal ones
\cite{Ratti:07,Fukushima:04plb} and thus the logarithmic divergence 
avoids an expectation value higher than one. This means that the 
logarithmic potential describes the dynamics of gluons more 
correctly and effectively. Keeping this in mind, we will be mainly
focusing on the discussion of the results in our calculation with 
the inclusion of the logarithmic Polyakov loop potential, though 
the curves of the calculation with the polynomial Polyakov loop 
potential will also be shown. The real physical effect of the 
Polyakov loop potential inclusion in the QMS model on mesonic
excitations, will become apparent when the results of our 
calculation in the PQMS models are compared with the corresponding 
results in the QMS model.


\subsection{ Meson Mass Variations }

\begin{figure*}[!tbp]
\subfigure[With $U_A(1)$ anomaly term]{
\label{fig:mini:fig2:a} 
\begin{minipage}[b]{0.4\linewidth}
\centering \includegraphics[width=\linewidth]{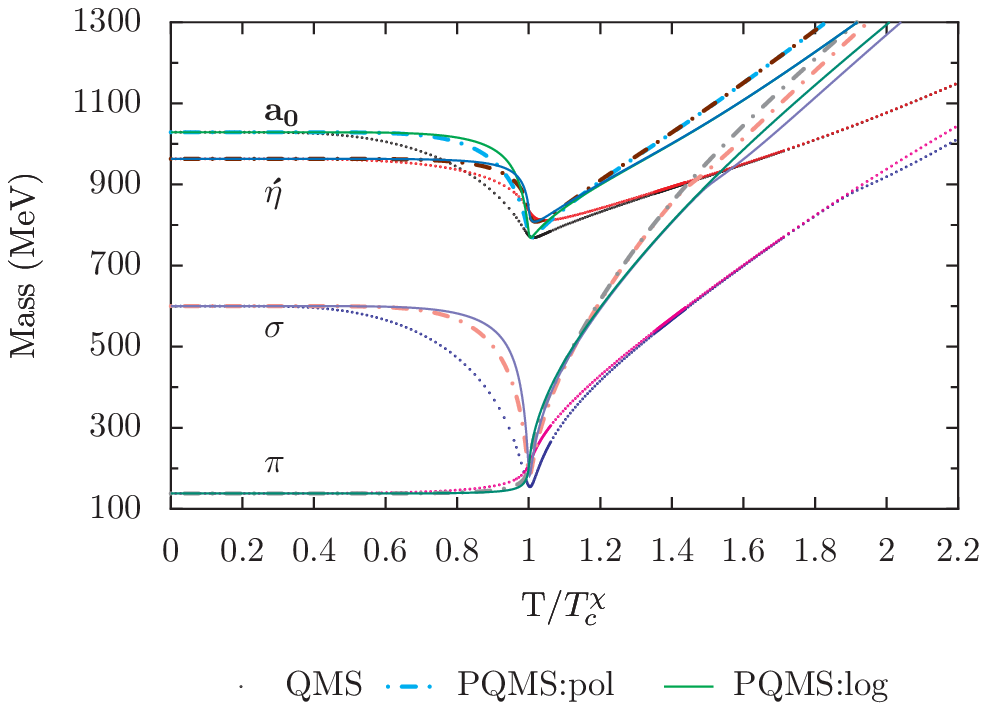}
\end{minipage}}%
\hspace{0.4in}
\subfigure[Without $U_A(1)$ anomaly term]{
\label{fig:mini:fig2:b} 
\begin{minipage}[b]{0.4\linewidth}
\centering \includegraphics[width=\linewidth]{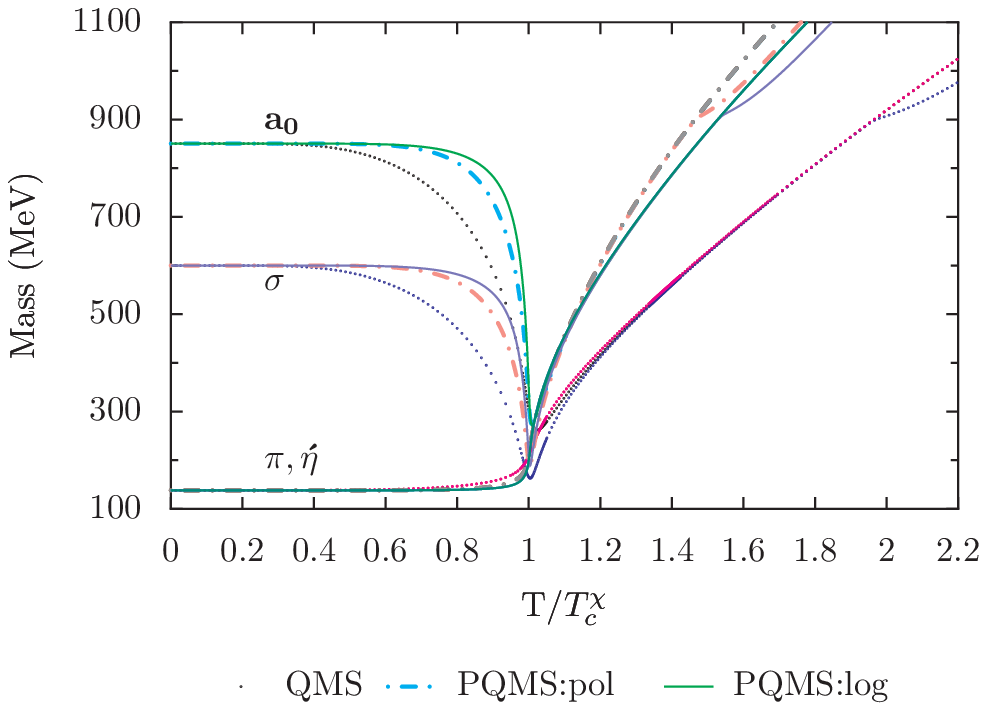}
\end{minipage}}
\caption{The mass variations of the chiral partners as functions of 
reduced temperature ($T/T_{c}^{\chi}$) at zero chemical potential
($\mu = 0$), in the 
presence of axial $U_A(1)$ breaking term, are plotted for 
($\sigma$, $\pi$) and ($a_0$, $\eta'$) in Fig.\ref{fig:mini:fig2:a} 
and the corresponding mass variations, in the absence of the $U_A(1)$ 
axial breaking term, are plotted in Fig.\ref{fig:mini:fig2:b}. The 
dotted line plots are the mass variations in the pure QMS model, dashed- 
dotted line plots represent the PQMS:pol model results, and the solid line 
plots are the mass variations in the PQMS:log model.}
\label{fig:mini:fig2} 
\end{figure*}

\begin{figure*}[!tbp]
\subfigure[With $U_A(1)$ anomaly term]{
\label{fig:mini:fig3:a} 
\begin{minipage}[b]{0.4\linewidth}
\centering \includegraphics[width=\linewidth]{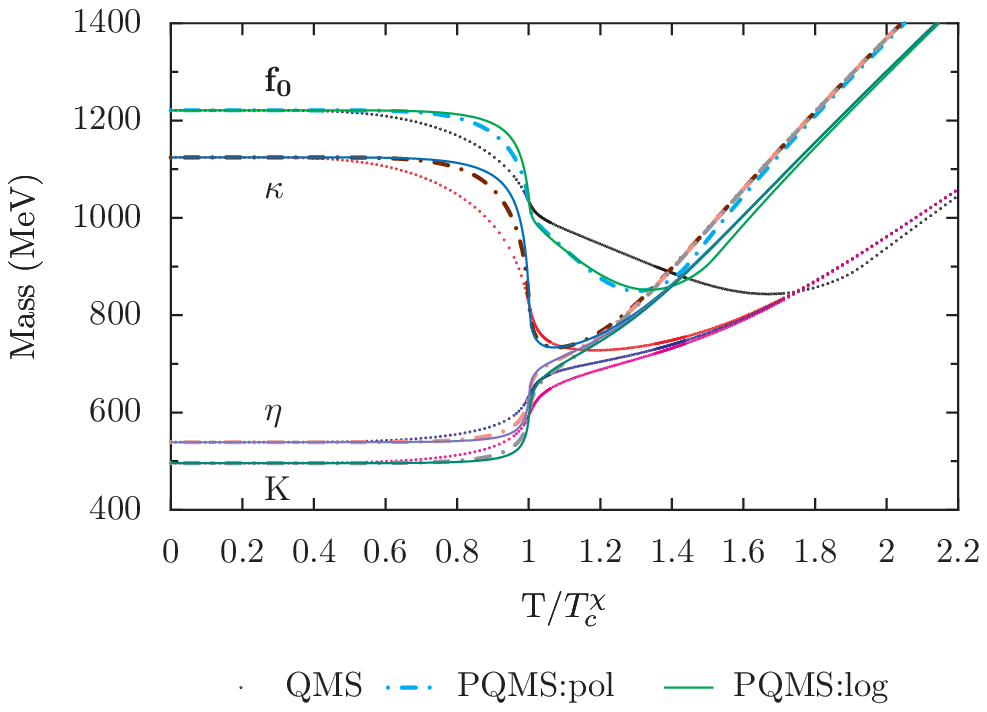}
\end{minipage}}%
\hspace{0.4in}
\subfigure[Without $U_A(1)$ anomaly term]{
\label{fig:mini:fig3:b} 
\begin{minipage}[b]{0.4\linewidth}
\centering \includegraphics[width=\linewidth]{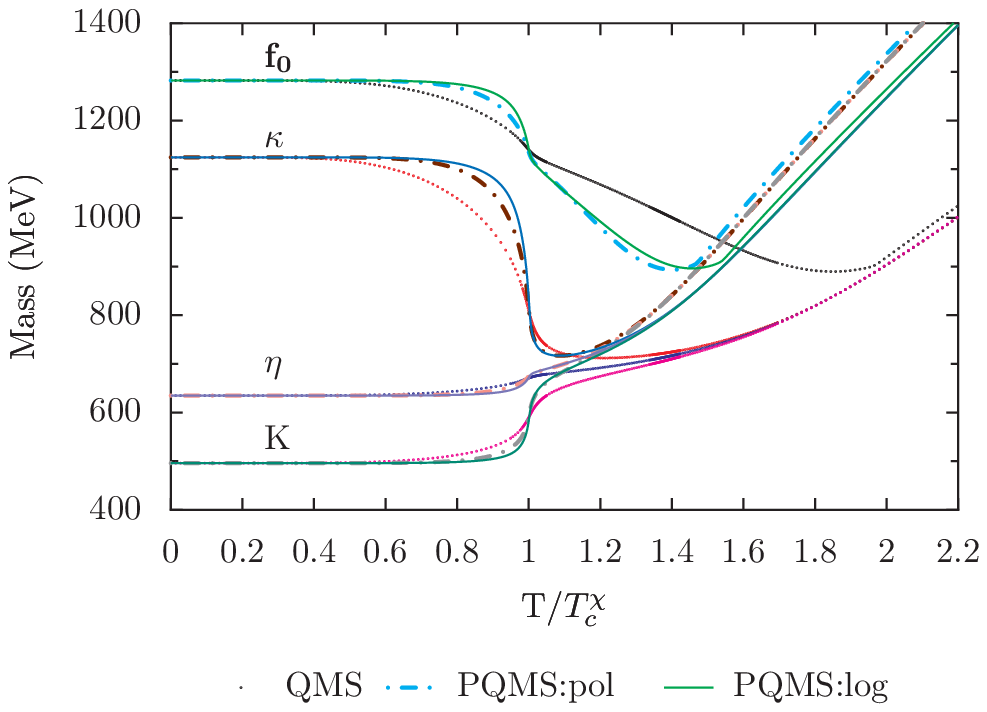}
\end{minipage}}
\caption{The mass variations of the chiral partners as functions of
reduced temperature ($T/T_{c}^{\chi}$) at zero chemical potential 
($\mu = 0$), in the presence of axial $U_A(1)$ breaking term, are 
shown for ($\eta$, $f_0$) and ($K$, $\kappa$) in Fig.\ref{fig:mini:fig3:a}, 
and the corresponding mass variations, in the absence of the $U_A(1)$
axial breaking term, are shown in Fig.\ref{fig:mini:fig3:b}. The dotted 
line plots are the mass variations in the pure QMS model, dashed-dotted line 
plots represent the PQMS:pol model results, and the solid line plots are 
the mass variations in the PQMS:log model. }
\label{fig:mini:fig3} 
\end{figure*}

We are calculating the masses of the scalar and pseudoscalar mesons
at finite temperature in the presence of the Polyakov loop potential
in the QMS model. We have collected 
the vacuum value of all the scalar and pseudoscalar meson masses in 
Table \ref{tab:mass}. The mass modifications calculated at finite 
temperature (Eq.\ref{eq:ftmass}) will be added to the 
vacuum masses of Table \ref{tab:mass}. The mass variations of the chiral 
partners as functions of reduced temperature, in the presence of 
axial $U_A(1)$ breaking term, are plotted for ($\sigma$, $\pi$) and 
($a_0$, $\eta'$) in Fig.\ref{fig:mini:fig2:a} and for ($\eta$, $f_0$) 
and ($K$, $\kappa$) in Fig.\ref{fig:mini:fig3:a}, while  the corresponding 
mass variations, in the absence of the $U_A(1)$ axial breaking term, 
are plotted in Fig.\ref{fig:mini:fig2:b} and Fig.\ref{fig:mini:fig3:b}. 
Further, since the focus of our investigation 
is the influence of the Polyakov loop on the effective restoration of 
symmetries, we will be comparing the mesonic observables below 
and above $T_{c}^{\chi}$.

In Fig.\ref{fig:mini:fig2:a}, the chiral partners ($\sigma$, $\pi$) 
and ($a_0$, $\eta'$) become mass degenerate in the close vicinity
of reduced temperature $T/T_c^{\chi}=1$. The masses of these particles
are dominated by the contribution from the nonstrange quarks and a rapid
crossover in the nonstrange sector (Fig.\ref{fig:fig1}) appears as 
sharper and faster mass degeneration in our calculation in the PQMS model.
Thus, the Polyakov loop inclusion in the QMS model makes a sharper mass 
degeneration as well as faster occurrence of chiral $SU_L(2) \times SU_R(2)$ 
symmetry restoration transition in the nonstrange sector.

In Fig.\ref{fig:mini:fig3:a}, the presence of the Polyakov loop potential
in the QMS model generates, the similar trend of sharper and faster 
mass degeneration in the masses of the chiral partners ($\eta$,$f_0$)
and ($K$, $\kappa$). Though the mass degeneration of chiral partners 
($K$, $\kappa$) with $\eta$ does not occur at $T/T_c^{\chi}=1$, it 
sets up early in the PQMS models at $T/T_c^{\chi}=1.3$, while it occurs 
at $T/T_c^{\chi}=1.5$ in the QMS model. In the PQMS models, the 
intersection point of the $f_0$ and $\eta$ masses, occurs early when
$T/T_c^{\chi}=1.4$, while in the QMS model it is found at $T/T_c^{\chi}=1.7$. 
This trend of mass degeneration reflects the effect of the Polyakov loop
potential on chiral symmetry restoration in the strange sector and it 
results due to sharper and stronger melting of the strange condensate
(Fig.\ref{fig:fig1}) in the influence of the Polyakov loop potential 
in the PQMS models.

The $U_A(1)$ breaking generates the mass gap between the two sets of
the chiral partners, ($\sigma$, $\pi$) and ($a_0$, $\eta'$), i.e. 
$m_{\pi} = m_{\sigma}$ $<$ $m_{a_0} = m_{\eta'}$ for $T/T_C^{\chi}>1$.
This mass gap results due to the opposite sign of the anomaly term 
($\sqrt{2}c\sigma_y$) in the scalar and pseudoscalar meson masses.
Hence, it will be reduced due to the melting 
of the strange order parameter $\sigma_y$ for the higher values of 
the reduced temperature $T/T_c^{\chi}>1$. Since the melting of the
strange condensate is stronger and sharper (Fig.\ref{fig:fig1}) in the 
PQMS model, the convergence in the masses of 
the two sets of chiral partners gets enhanced in these 
calculations. Thus the inclusion of the Polyakov loop potential in the 
QMS model also effects an early set up of $U_A(1)$ restoration trend 
on the reduced temperature scale. 

Now we discuss the variations in the masses of chiral partners when 
the explicit $U_A(1)$ symmetry breaking term has been taken as zero 
(c=0).  We notice in Fig.\ref{fig:mini:fig2:b} and Fig.\ref{fig:mini:fig3:b},
again, the same sharper and faster trend of mass degeneration that 
we identify as the effect generated by the inclusion of the Polyakov 
loop potential. The $\eta'$ meson degenerates with the pion in vacuum 
and stays the same for all temperatures in Fig.\ref{fig:mini:fig2:b}
due to the absence of the anomaly term. Further the mass gap between
the chiral partners ($\sigma$, $\pi$) and ($a_0$, $\eta'$) becomes
zero and all four of the mesons become degenerate at $T/T_c^{\chi}=1.0$.
The $T/T_c^{\chi}$ numerical value, where the $K$, $\kappa$, and $\eta$
masses degenerate in different models, is not influenced by the 
$U_A(1)$ anomaly as expected since the nonstrange condensate does not 
have any anomaly dependence. Further,in Fig.\ref{fig:mini:fig3:b}, 
the intersection point of the $f_0$ and $\eta$ masses in the PQMS 
models, is obtained when $T/T_c^{\chi}$ is around 1.6  while in 
the QMS model, this intersection point is found around 
$T/T_c^{\chi}=2.0$. We are also obtaining the mild anomaly 
dependence of the intersection point of the $f_0$ and $\eta$ in all 
the models. Here, we note that the mass of $f_0$ in vacuum 
increases by about 60 MeV in the absence of anomaly.

The temperature variations of meson masses, in general, result due to
the interplay of the bosonic thermal contributions (decreasing 
the meson masses) and fermionic quark contributions (increasing
the meson masses). Quark contributions which are negligible at small
temperatures, dominate the mesonic contributions for high temperatures,
and this generates a rising trend in meson masses, which ultimately 
leads to the mass degeneration of the chiral partners
\cite{Schaefer:09}. In the PQMS models, the one
quark and two quark fermionic contributions are suppressed due to the
presence of the Polyakov loop potential. Since the chiral phase transition is
driven by the fermionic contributions, chiral restoration gets delayed
due to the delay in the deconfinement transition and because of this we 
get higher value of the pseudocritical temperature $T_{c}^{\chi}$ in the
Polyakov loop augmented quark meson linear sigma model. The higher value of the
$T_{c}^{\chi}$ makes the ratio $T/T_{c}^{\chi}$ small. Hence, in 
comparison to the QMS model, the mass degeneration trend among the 
chiral partners, in general, sets up early in the PQMS models on the 
reduced temperature scale.

The variation of meson masses with the
polynomial Polyakov loop potential are similar though a little less
sharp than the mass calculations with the logarithmic potential. 
The difference appears 
mainly because of difference in the Polyakov loop expectation
value $\langle\Phi\rangle$ with these two potentials. 
The calculations with the polynomial Polyakov loop potential make sense only
for $T < 2T_c^{\chi}$.

The mass variation of scalar $\sigma$ 
and $f_0$ show kink around $T/T_c^{\chi}=1.8$ in the QMS model
while it is seen around $T/T_c^{\chi}=1.4$ in the PQMS models. The 
kink generation results because the meson masses seem to
interchange their identities for higher values of the reduced 
temperature \cite{Schaefer:09}. In order to have a proper perspective
of the kink behavior in the curves, one has to study and analyze 
the scalar and pseudoscalar meson mixing angles.

\subsection{Meson Mixing Angle Variations}

\begin{figure*}[!t]
\centering
\hspace{-0.3cm}\includegraphics[width=0.8\linewidth]{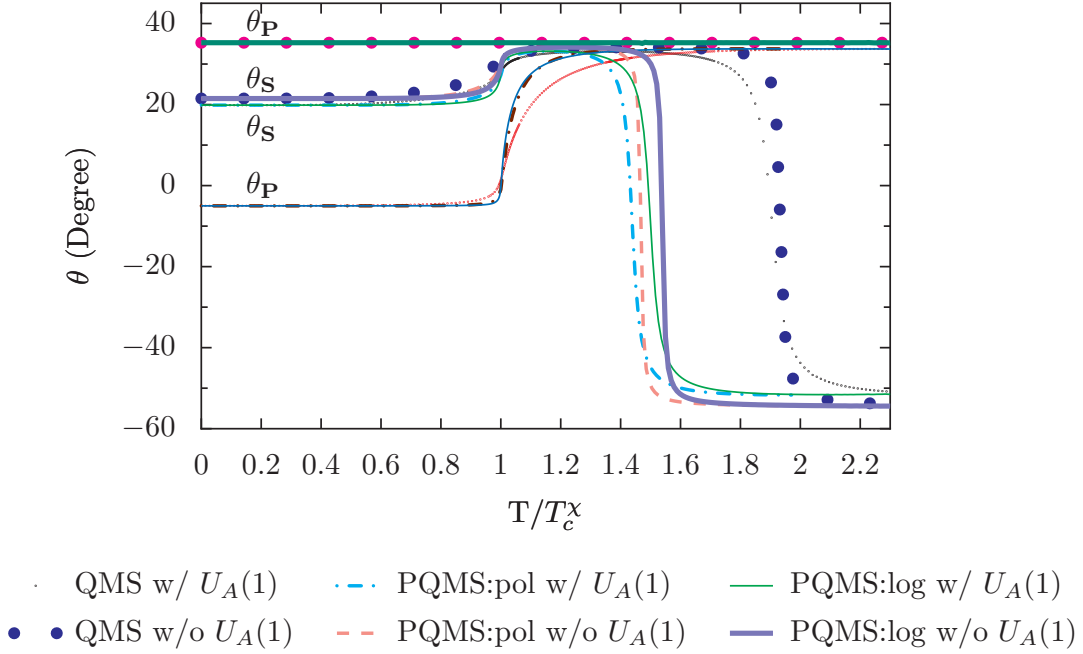}
\caption{The scalar $\theta_S$ and pseudoscalar $\theta_P$ mixing angle 
variations with respect to the reduced temperature ($T/T_{c}^{\chi}$) 
at zero chemical potential ($\mu = 0$) are plotted. We have given the 
plots for the QMS, PQMS:log, and PQMS:pol models considering the cases in the 
presence as well as absence of the axial $U_A(1)$ explicit symmetry breaking 
term. The dotted lines show the result with anomaly in the QMS model while
the solid big dot lines show the result without anomaly. In the PQMS:log 
model, the thick solid lines represent the variations without anomaly while 
thin solid lines show the result with anomaly. The dashed-dotted lines are
the variations with anomaly in the PQMS:pol model, while the dashed lines are 
the corresponding results in the absence of anomaly.}
\label{fig:thplt}
\end{figure*}

The analysis of axial $U_A(1)$ restoration pattern identification 
will become complete, only after studying the variation of scalar 
$\theta_S$ and pseudoscalar $\theta_P$ mixing angles on the relative 
temperature scale in Fig.\ref{fig:thplt} considering the cases in 
the presence as well as the absence of the axial $U_A(1)$ 
explicit symmetry breaking term. The 
anomaly term has a strong effect in the pseudoscalar sector in the
broken phase for $T/T_c^{\chi}<1$ while no effect of anomaly is found 
in the scalar sector. The nonstrange and strange quark mixing is 
strong, at T=0 one gets $\theta_P=-5^{\circ}$,  which remains almost
constant in the chiral broken phase. In the vicinity of
$T/T_c^{\chi}=1$, the $\theta_P$ variations start approaching the 
ideal mixing angle $\theta_P \rightarrow \arctan{\frac{1}{\sqrt{2}}} \sim 
35^{\circ}$ , the corresponding  $\Phi_P = 90^\circ$. Here, $\Phi_P$ 
is the pseudoscalar mixing angle in the strange nonstrange basis 
(see Ref.\cite{Schaefer:09} for details). The smooth approach towards 
the ideal mixing in the QMS model, becomes sharper and faster in the PQMS 
model calculations due to the influence of the Polyakov loop potential.
Further, the ideal mixing is achieved earlier on the reduced 
temperature scale in the PQMS models. In the absence of axial 
$U_A(1)$ anomaly, the pseudoscalar mixing angle remains ideal
$\theta_P=35^\circ$ everywhere on the reduced temperature scale. 
 
\begin{figure*}[!tbp]
\subfigure[With $U_A(1)$ anomaly term]{
\label{fig:mini:crs:eta} 
\begin{minipage}[b]{0.45\linewidth}
\centering \includegraphics[width=0.9\linewidth]{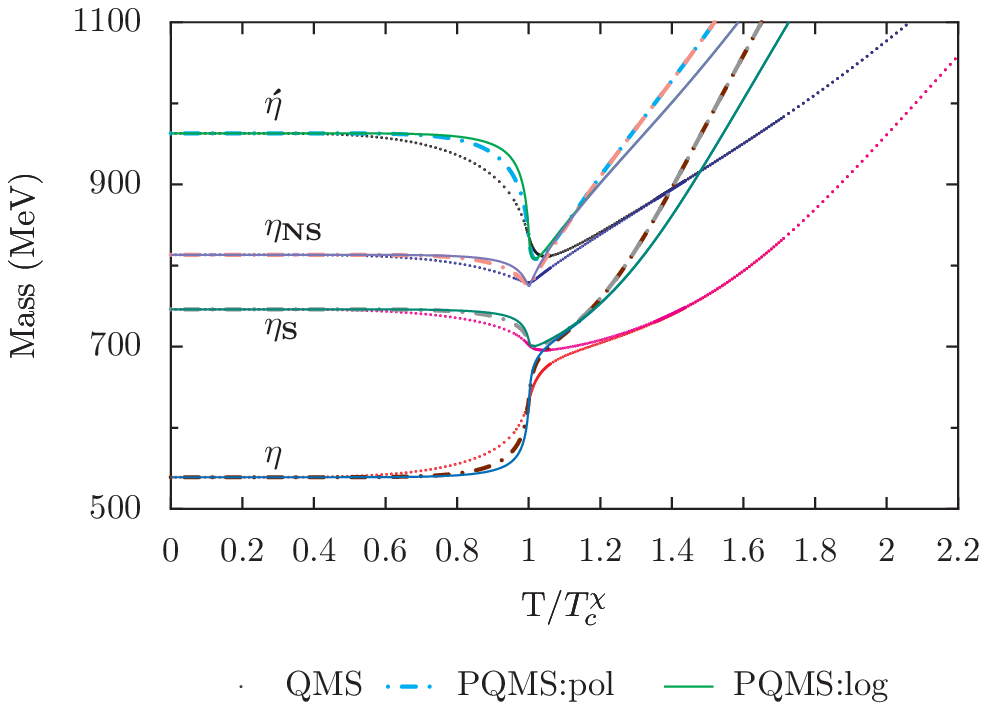}
\end{minipage}}%
\hfill
\subfigure[With $U_A(1)$ anomaly term.]{
\label{fig:mini:crs:sig} 
\begin{minipage}[b]{0.45\linewidth}
\centering \includegraphics[width=0.96\linewidth]{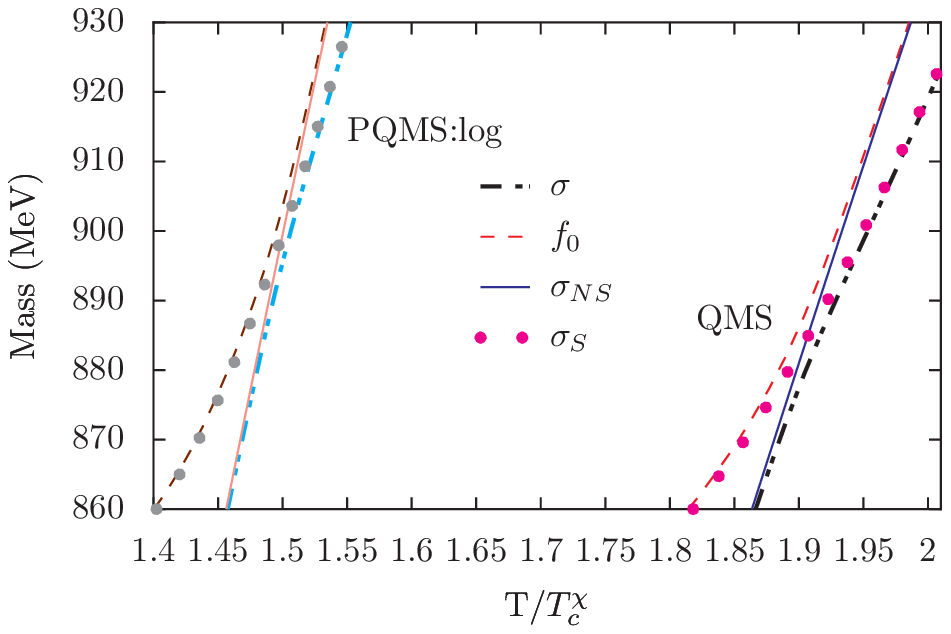}
\end{minipage}}
\caption{Figure\ref{fig:mini:crs:eta} shows 
the mass variations for the physical $\eta$, $\eta'$ and the 
nonstrange-strange $\eta_{NS}$, $\eta_{S}$ complex, on the reduced 
temperature scale ($T/T_{c}^{\chi}$) at zero chemical potential ($\mu = 0$).
The masses of the physical $\sigma$ and $f_0$ anticross and the 
nonstrange-strange $\sigma_{NS} - \sigma_{S}$ system masses cross in
Figure\ref{fig:mini:crs:sig}.}
\label{fig:mini:crs} 
\end{figure*}

The $\eta$ and $\eta'$ mesons become a purely strange $\eta_S$ 
and nonstrange $\eta_{NS}$ quark system as a consequence of the 
ideal pseudoscalar mixing, which gets fully achieved at higher 
values of the reduced temperature. In order to show this, we have
plotted in Fig.\ref{fig:mini:crs:eta} the mass variations for the
physical $\eta$, $\eta'$ and the nonstrange-strange $\eta_{NS}$, 
$\eta_S$ complex. Mass formulae $m_{\eta_{NS}}$ and $m_{\eta_S}$ are 
given in Table \ref{tab:mass}. Again, the smooth mass convergence 
trend, of the pure QMS model in $m_{\eta'} \rightarrow m_{\eta_{NS}}$ and 
$m_{\eta} \rightarrow m_{\eta_S}$ approach, becomes sharper and faster 
around $T/T_c^{\chi} = 1$ in the influence of the Polyakov loop potential 
in the QMS model. The exact $m_{\eta'} \rightarrow m_{\eta_{NS}}$ and 
$m_{\eta} \rightarrow m_{\eta_S}$ 
mass convergence in the PQMS models, occurs closer to the value 
$T/T_c^{\chi}=1$.

In Fig.\ref{fig:thplt} for $m_{\sigma}$ = 600 MeV at T = 0 
scalar mixing angle $\theta_S \sim 19.9^{\circ}$ in the
presence of anomaly, while $\theta_S \sim 21.5^{\circ}$ in 
the absence of anomaly. 
The $\theta_S$ around $T/T_c^{\chi}=1$ grows 
to its ideal value but for higher temperatures on the reduced 
temperature scale, in the chiral symmetric phase, the scalar 
mixing angle drops down to $\theta_S \sim -51^{\circ}$ and
$\theta_S \sim -54^{\circ}$ in the respective cases considered 
with and without anomaly. In the 
presence of the $U_A(1)$ symmetry breaking term, this drop happens 
in the QMS model around $T/T_c^{\chi} \sim 1.9$ and due to the
effect of the Polyakov loop potential the similar drop occurs earlier
for $T/T_c^{\chi} \sim 1.5$ in the PQMS model. In the close 
vicinity of these reduced temperatures, the masses of the physical
$\sigma$ and $f_0$ anticross and the nonstrange - strange 
($\sigma_{NS} - \sigma_S$) system masses cross as shown in 
Fig\ref{fig:mini:crs:sig}. It means that after
anticrossing the physical $\sigma$ becomes identical with the pure 
strange quark system $\sigma_S$, while the physical $f_0$ becomes 
degenerate with the pure nonstrange quark system $\sigma_{NS}$. 
A similar drop for the calculations without anomaly happens at a 
little higher value on the reduced temperature scale in respective
models.


\section{Summary and discussion}
\label{sec:smry}

We have calculated the meson masses and mixing angles for the
scalar and pseudoscalar sector in the framework of the generalized 
2+1 flavor PQMS model. We have used two different forms of the 
effective Polyakov loop potential for the calculation, namely, 
the polynomial potential and logarithmic potential. In order to 
investigate the influence of Polyakov loop potential on chiral 
symmetry restoration, these calculations have been compared with 
the corresponding results in the QMS model.

The temperature dependence of nonstrange, strange condensates and 
the Polyakov loop field $\Phi$ at zero chemical potential has been 
calculated from the gap equation in the QMS and PQMS models. 
Comparison of pseudocritical temperatures calculated from the 
inflection points of these order parameters indicates, that the 
chiral transition gets shifted to the higher temperatures as a result
of the inclusion of the Polyakov loop in the QMS model. We further
observe that the variation of the nonstrange condensate in the 
$T/T_{c}^{\chi}$ = 0.8 to 1.2  range becomes quite sharp due to the
effect of the Polyakov loop potential in our calculation in PQMS 
models. We infer from the curves in the PQMS models that the inclusion 
of the Polyakov loop potential in the QMS model together with the
presence of axial anomaly, triggers  an early and significant melting 
of the strange condensate. The interesting physical consequences of 
the earlier melting of the strange condensate are an early emergence 
of mass degeneration trend in the masses of the chiral partners 
($K$, $\kappa$) and ($\eta$, $f_0$) and an early setting up of a 
$U_A(1)$ restoration trend.

The mass degeneration of chiral partners ($\sigma$, $\pi$) and 
($a_0$, $\eta'$) in the close vicinity of $T/T_{c}^{\chi}=1.0$ 
becomes sharper and faster in our calculations in the PQMS model. 
This sharpening of the mass variations in the small neighborhood 
of $T/T_c^{\chi}=1$ results due to the stronger and sharper melting 
of the nonstrange condensate triggered by the presence of the 
Polyakov loop potential in the QMS model. Thus, we can corroborate 
also from the behavior of the chiral partners that the net effect of 
the Polyakov loop inclusion in the QMS model, is to make a sharper 
occurrence of the chiral $SU(2)_L \times SU(2)_R$ symmetry restoration
transition in the nonstrange sector. Further, the mass degeneration 
of chiral partners ($K$, $\kappa$) with $\eta$ does not occur when 
the value of the reduced temperature is equal to one, it sets up 
early in the PQMS models at $T/T_c^{\chi}=1.3$, while it occurs at 
$T/T_c^{\chi}=1.5$ in the QMS model. In the PQMS models, the 
intersection point of the $f_0$ and $\eta$ masses, occurs early 
when the reduced temperature $T/T_c^{\chi}=1.3$, while in the 
pure QMS model this intersection point is found at $T/T_c^{\chi}=1.7$. 
This trend of mass degeneration emerges, again as a result of the
sharper and stronger melting of the strange condensate in the 
influence of the Polyakov loop potential in the PQMS models.

The $U_A(1)$ breaking anomaly effect that leads to the mass gap 
between the two sets of the chiral partners,($\sigma$, $\pi$) and 
($a_0$, $\eta'$) i.e. $m_{\pi} = m_{\sigma}$ $<$ $m_{a_0}=m_{\eta'}$ 
for $T/T_C^{\chi}>1$, is proportional to the strange condensate 
$\sigma_y$. Since the melting of the strange condensate is stronger
and sharper in the PQMS models, the convergence in the masses of
the two sets of chiral partners will be enhanced in these 
calculations. Thus, the inclusion of the Polyakov loop potential
in the PQMS models also effects an early set up of the $U_A(1)$ 
restoration trend on the reduced temperature scale.

The smooth approach of the pseudoscalar mixing angle $\theta_P$
towards the ideal mixing in the QMS model, becomes sharper and faster
in the PQMS models due to the influence of the Polyakov loop potential. 
Further, in comparison to QMS model results, the ideal mixing on 
the reduced temperature scale is achieved earlier in the PQMS models.
The $\theta_S$ around $T/T_c^{\chi}=1$ grows to its ideal value but 
for higher temperatures on the reduced temperature scale, in the 
chirally symmetric phase, the scalar mixing angle drops down to 
$\theta_S \sim -51^{\circ}$. In the presence of $U_A(1)$ symmetry
breaking term, this drop happens in the QMS model for 
$T/T_c^{\chi} \sim 1.85$ and in the PQMS:log model, the similar 
drop occurs for $T/T_c^{\chi} \sim 1.5$. In the close vicinity of 
these reduced temperatures, the masses of the physical 
$\sigma$ and $f_0$ anticross and the nonstrange-strange 
$\sigma_{NS} - \sigma_S$ system masses cross.

%
\begin{acknowledgments}
We are very thankful to Ajit Mohan Srivastava, Tamal K. Mukherjee and 
Neelima Agarwal for valuable discussions. We acknowledge the support 
of the Department of Atomic Energy- Board of Research in Nuclear 
Sciences (DAE-BRNS), India, under the research grant No. 2008/37/13/BRNS.
We also acknowledge the computational support of the computing facility
which has been developed by the Nuclear Particle Physics group of the Physics
Department, Allahabad University under the Center of Advanced Studies(CAS)
funding of UGC India. 
\end{acknowledgments}


\end{document}